# New hybrid organic-inorganic ferrophotovoltaic perovskites nanoparticles with high voltage for indoor and IoT applications

Rémi Ndioukane[a,*], Fanta Baldé[a], Ndéye C. Y. Fall[a], Diouma Kobor[a], Laurence Motte[b]

[a]*Laboratoire de Chimie et de Physique des Matériaux (LCPM), Assane Seck University of Ziguinchor, Senegal*

[b]*Laboratory for Vascular Translational Science (LVTS), Paris 13 University, France*

**Abstract**

The ideal band gap for a photovoltaic active layer for the solar spectrum is around 1.3 eV. However oxides with such values are rare. One of the most studied oxides to date as a photovoltaic active layer is the cuprous oxide $Cu_2O$. Its band gap is around 2.1 eV and is therefore not ideal for the solar spectrum. Power Conversion Efficiency generally do not exceed 4%. In this paper we propose to study an emerging type of solar cell that is based on ferroelectricity. In this type of solar cell, a p-n junction is not necessarily required, unlike conventional solar cells. Interesting conversion efficiencies are beginning to be obtained with this type of cell, however the mechanisms are still not well understood and several material and engineering challenges must be addressed. The objective of this paper is to initiate an innovative photovoltaic technology based on novel inorganic with suitable bandgap widths and organic materials (biopolymer). These oxides are more stables. We synthesized ferroelectric materials that absorb a large part of the solar spectrum with reduced bandgap widths. PZN-4.5PT nanoparticles were dispersed in a biopolymer matrix. Hybrid thin films with these inorganic nanoparticles embedded in a biopolymer have been successfully fabricated by spin coating on ITO substrate. Structural, morphological and electrical properties were investigated. The best Power Conversion Efficiencies measure under a light LED illumination of 3550 lux are respectively 21.83 % and 31.62 % for 15 and 30 min light exposition with an open-circuit voltage of 5.17 and 5.86 V.

## 1. Introduction

The photovoltaic effect is used to directly harvest solar energy by converting the incident photons into flowing free charge carriers and thus produce electrical current. Y. M. Fridkin proposed mechanism and related phenomena for possible conversion of light to electricity using bulk

Corresponding Author
*Email adress:* r.ndioukane1532@zig.univ.sn (Rémi Ndioukane)

ferroelectrics, i.e. "photo-ferroelectrics" [1]. In the subsequent decade further development has been taken place on ferroelectric photovoltaic (PV) materials and devices [2] mostly centered, however, around the understanding of basic physics. Photo-ferroelectrics are a particular class of materials where phenomena of the ferroelectric effect and the photoelectric or photovoltaic effect are intimately linked. These two effects are somewhat incompatible in nature since high quality ferroelectrics must be electrical insulators in order to withstand large applied voltages or fields, whereas photovoltaic devices need high open-circuit voltage and short-circuit current. A big advantage of photo-ferroelectrics compared to conventional silicon or organic photovoltaic devices is their direction: polarization dictates the direction of photocurrents. The photoferroic systems develop a complex relation between the ferroelectric phase stability (including domain size and distribution) and photoresponse; these photo-generated charge carriers (electrons and holes) in relatively high concentration affect the free energy close to ferroelectric transition (Curie point). The presence of charge carriers may cause structural deformation where the unit cell volume gets affected by generation of charge carriers during the phase transition [3]. Among the ferroelectrics families lead zirconate titanate $PbZr_xTi_{1-x}O_3$ (PZT) found a special place due to its optimal properties, such as large fatigue-free polarization, piezoelectricity, and good electro-optic effect. These properties are related to polarization behavior under the influence of external electric field [4]. Since the discovery of PZT, extensive research has been carried out on substitution of various types of di- and tri-valence cations on lead and zirconia/titanate sites. Some of these substitutions were very successful and led to development of device quality materials with improved optical, ferroelectric, dielectric and leakage properties [5].

However, with the advent of sophisticated thin film fabrication techniques, ferroelectric photovoltaic effects and low bandgap ferroelectric materials received great attention due to strong coupling of polarization and light. Ferroelectric oxides are also stable in a wide range of mechanical, chemical and thermal conditions and can be fabricated using low-cost methods such as sol–gel thin film deposition and sputtering [6].

The objective of this paper is to initiate an innovative photovoltaic technology based on novel inorganic with suitable bandgap widths and organic materials (biopolymer).

## 2. Experimental procedure

To produce thin layers of perovskite nanoparticles, the perovskite crystal was ground dry until a very fine powder was obtained. For the dispersion of perovskite nanoparticles, biopolymer was used as a matrix. Obtaining the biopolymer required extraction from the bark of a tree. These barks were then soaked in water for at least 24 hours and then a viscous solution was recovered and were act as a gel (matrix). PZN-PT perovskite nanoparticles were dispersed inside the biopolymer matrix and a homogeneous solution containing ferroelectric oxide perovskite nanoparticles was obtained after 30 minutes of stirring under a magnetic stirrer. After depositing this layer of perovskite nanoparticles by spin coating, a layer of perylene was then deposited. This layer facilitates the absorption of incident light. It is used to transport the holes. Thus, the deposition of the various thin layers ITO/TiO2/np-PZN4.5PT+Biop/pery, leads to the creation of a hybrid solar cell based on nanoparticles of inorganic perovskites of ferroelectric oxide. For charge collection, a counter electrode was deposited on a sensitizing layer which acts as a charge collector.

Scanning electron microscopy was used to characterize the surface morphology. The current-voltage characteristics are measured at room temperature using the Keithley I-V source meter 2612B device which is coupled with a computer under the TSP Express software. Measurements were carried out on solar cells with a surface area of 3.2 cm$^2$ under a light LED of 3550 lux.

## 3. Results and discussion

Figure 1 shows SEM images with a slightly nonhomogeneous distribution of nanoparticles inside the device, which certainly explain the low current density. One can see the accumulation of perovskites nanoparticles on the biopolymer surface. This reduces a better extraction of charges while increasing recombination kinetics at the interface.

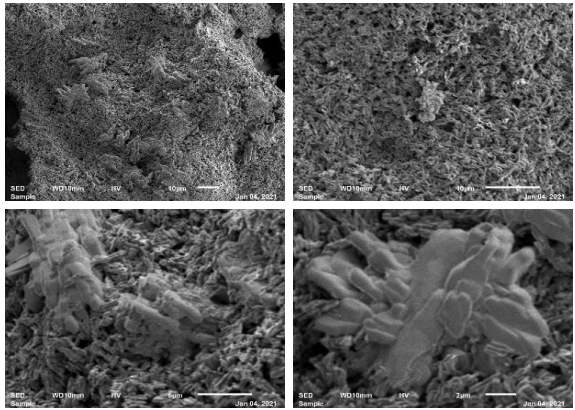

Figure 1: SEM images of nanoparticles on the biopolymer surface

The long exposition on light gave us different I-V characteristics which increase the energy yield during the time (Figure 2). After 15 min the PCE is 21.83 % and 31.62 % for 30 min light exposition. The open circuit voltage ($V_{oc}$) value is 5.17 V at 15 min and 5.86 V for 30 min. The short-circuit current density ($J_{sc}$) is constant ($1.25 \times 10^{-4}$ A/cm$^2$). The fill factor (FF) is 0.27 and 0.34 after respectively 15 and 30 min light exposition. The $V_{oc}$ values are very high compared to those of $J_{sc}$. Indeed, Ilya Grinberg et al. showed that these materials could reach values above 10 V [7]. The calculation of the maximum power $P_{max}$ gave 0.43 and 0.63 mW/cm$^2$ respectively for 15 and 30 min under 3550 lux. This result shows a promising photovoltaic property for indoor application.

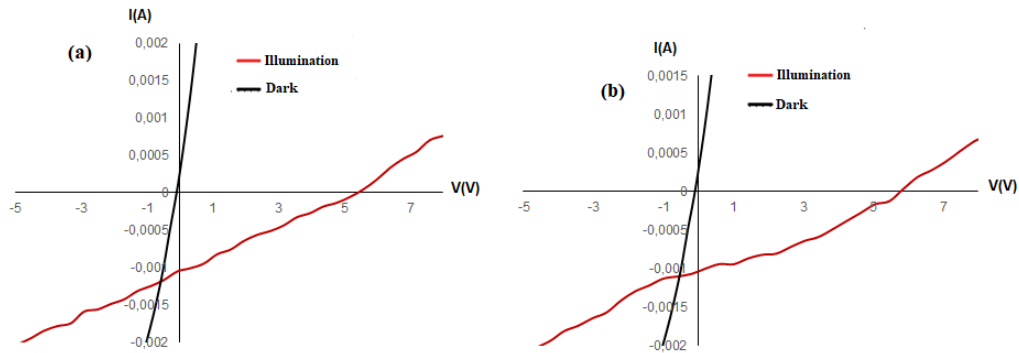

Figure 2: I-V characterization under dark and light illumination at a) 15 min and b) 30 min

This type of cell is one of the best cells in terms of conversion efficiency and voltage.
Different measurements were made to determine the electrical properties. We obtained different values of $V_{oc}$ and $I_{sc}$ as a function of the time of exposure to light (see Figure 3). In figure 3, We were able to show the variation of $V_{oc}(t)$, thus allowing us to see that the $V_{oc}$ increases over time with a max value of 12.72 V. This is due to the fact that ferroelectric perovskites have the ability to accumulate electrical charges on the surface of the sample thus can increase the $V_{oc}$ over time of exposure to light. As for the $I_{sc}$, no variation is noted over time; its value is 0.5 mA. We deduce that long exposure to light from a solar cell containing ferroelectric perovskites increase the $V_{oc}$ but has no effect on the short circuit current of the device. The efficiency is proportional to $V_{oc}$ and $I_{sc}$, will have a direct impact.

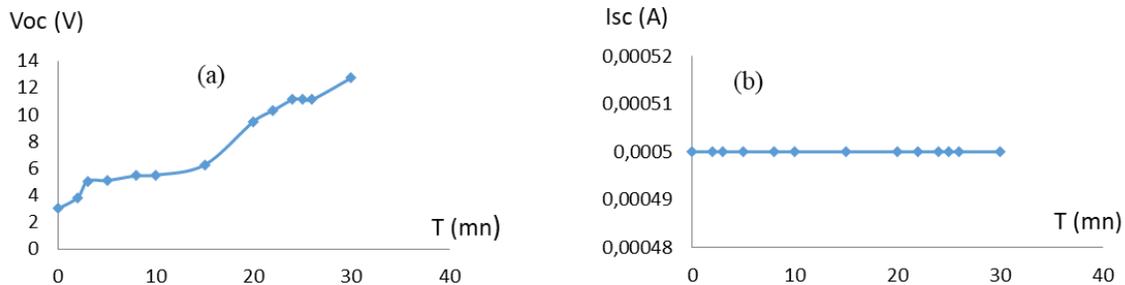

Figure 3: variation as a function of time of a) $V_{oc}$ and b) $I_{sc}$

## 4. Conclusion

In this study, we can note that long exposure to low power light from a solar cell containing ferroelectric perovskites increase the open circuit voltage but has no effect on the short circuit current of the device. The efficiency is proportional to $V_{oc}$ and $I_{sc}$, will have a direct impact. Pmax shows a promising photovoltaic property for indoor application.
Investigations are in progress particularly for the stability in time and the temperature effect.

**Acknowledgments**

This work is supported by the Senegalese Ministry of High Education